\documentclass[aps, aip, pra, prl, rsi, preprint, reprint, twocolumn, showpacs, preprintnumbers, amsmath, amssymb]{revtex4-1}

\usepackage{graphicx}
\usepackage{epstopdf} 
\usepackage{hyperref}
\usepackage{fancyhdr}

\begin{document}

\title{Simple Laser Stabilization to the Strontium $^{88}$Sr Transition at 707~nm}

\author{Matthew A. Norcia}
\author{James K. Thompson}
\affiliation{JILA, National Institute of Standards and Technology, and Department of Physics, University of Colorado, Boulder, Colorado 80309-0440, USA}
\email[]{matthew.norcia@colorado.edu}

\begin{abstract}
We describe frequency stabilization of a laser at 707.202~nm wavelength using FM spectroscopy in a hollow cathode lamp.  The laser is stabilized to the $^{88}$Sr metastable $^3\mathrm{P}_2$ to $^3\mathrm{S}_1$ optical transition.  The stabilized laser is utilized for laser-cooling and trapping of strontium atoms.  We also briefly describe the performance of a polarization spectroscopy lock of our blue MOT laser at 461~nm derived simultaneously from the same hollow cathode lamp.
\end{abstract}

\pacs{}

\maketitle

Laser cooled and trapped strontium and other alkaline-earth like atoms offer a rich internal structure that can be readily manipulated using laser light.  The many metastable excited states have enabled a broad range of new physics explorations including the most accurate and precise clocks \cite{NISTYb2009, Ye2014, ushijima2015cryogenic}, studies of SU(N) symmetric scattering interactions \cite{cazalilla2009, gorshkov2010}, and the engineering of synthetic gauge fields for quantum simulations with cold atoms \cite{jaksch2003}.  It may be possible to use strontium atoms as the gain medium for a new class of superradiant lasers with frequency linewidths of order 1~mHz \cite{Meiser09, Bohnet12a}.


In order to fully exploit strontium's rich internal structure, one must stabilize the frequency of lasers used to couple the internal levels, shown in Figure 1a.  Here we describe the stabilization of a grating laser to the metastable $^3\mathrm{P}_2$ to $^3\mathrm{S}_1$ optical transition in $^{88}$Sr.  The described laser system has already been applied in recent experiments in cavity-QED on a narrow forbidden transition \cite{Norcia2015a}. In the future, it will be applied to a superradiant laser to provide both optical pumping and continuous polarization gradient and Raman sideband cooling of the strontium atoms that will constitute the laser's gain medium \cite{Meiser09}.

Spectroscopic stabilization techniques for other wavelengths involved in the manipulation of strontium are now well established.  For transitions between the $^1\mathrm{S}_0$ electronic ground state and the $^3\mathrm{P}_1$ and $^1\mathrm{P}_1$ exited states, a heat-pipe can be used to form a vapor with sufficient optical depth for robust spectroscopy \cite{Katori04}. For stablization to transitions between metastable excited and higher-excited states, the thermal occupation of the metastable state in a heat-pipe provides insufficient optical depth for spectroscopy.   

Here we employ a hollow-cathode lamp in which collisions generate sufficient population  in the metastable $^3\mathrm{P}_2$ state for spectroscopy. Such a hollow-cathode lamp has been used to stabilize a laser to the $^1\mathrm{S}_0$ to $^1\mathrm{P}_1$ transition \cite{Yoshio13}. To our knowledge, however, this paper represents the first demonstration of the stabilization of a laser directly to an atomic transition in strontium that does not involve the electronic ground state.  We also briefly characterize how the same hollow cathode lamp is used to simultaneously derive a polarization spectroscopy signal for stabilizing the blue-MOT laser at 461~nm.

The 707~nm spectroscopy signal is used to frequency stabilize a grating laser that is configured in Littrow configuration. The laser consists of an AR coated diode  (Sacher Lasertechnik SAL-0705-020) and an 1800 lines/mm grating (Edmund Optics NT43-775) at approximately 2 cm distance from the diode.  A collimating lens (Thorlabs C330TME-B, f=3.1mm, NA=0.68) is  positioned to minimize the required lasing threshold current to 32.5~mA.   A fraction of the laser output is sent to the spectroscopy setup by coupling into a polarization maintaining fiber with polarizing beam cubes on the input and output of the fiber to reduce slow polarization drifts in the fiber.

A see-through hollow cathode lamp (Hamamatsu L2783-38NE-SR with an EMCO HC2012 HV supply, current set to 20~mA) is used to create a gas of strontium atoms, some fraction of which are prepared by the electron-collision process in the metastable excited state $^3\mathrm{P}_2$, which at room temperature has a measured lifetime of around 500~seconds \cite{yasuda2004lifetime}.  The fiber output is focused by a collimating  lens (Thorlabs C110TME-B, f=6.24 mm, NA=0.4) to generate counter propagating probe and pump beams with probe and pump waist sizes at the hollow cathode lamp $w_{pr} = 160~\mu$m and $w_{pu}= 220~\mu$m respectively.  Both beams are horizontally polarized.  The probe and pump have 0.20~mW and 2.3~mW of power respectively as they pass though the cathode.  Approximately 0.08~mW of the probe light is detected using a Hamamatsu S5972 photodiode (quantum efficiency 84\%) AC coupled to an AD8015  7~k$\Omega$ transimpedance amplifier, with a 6.7~k$\Omega$ DC monitor path.

\begin{figure*}
\includegraphics[width=6.5in]{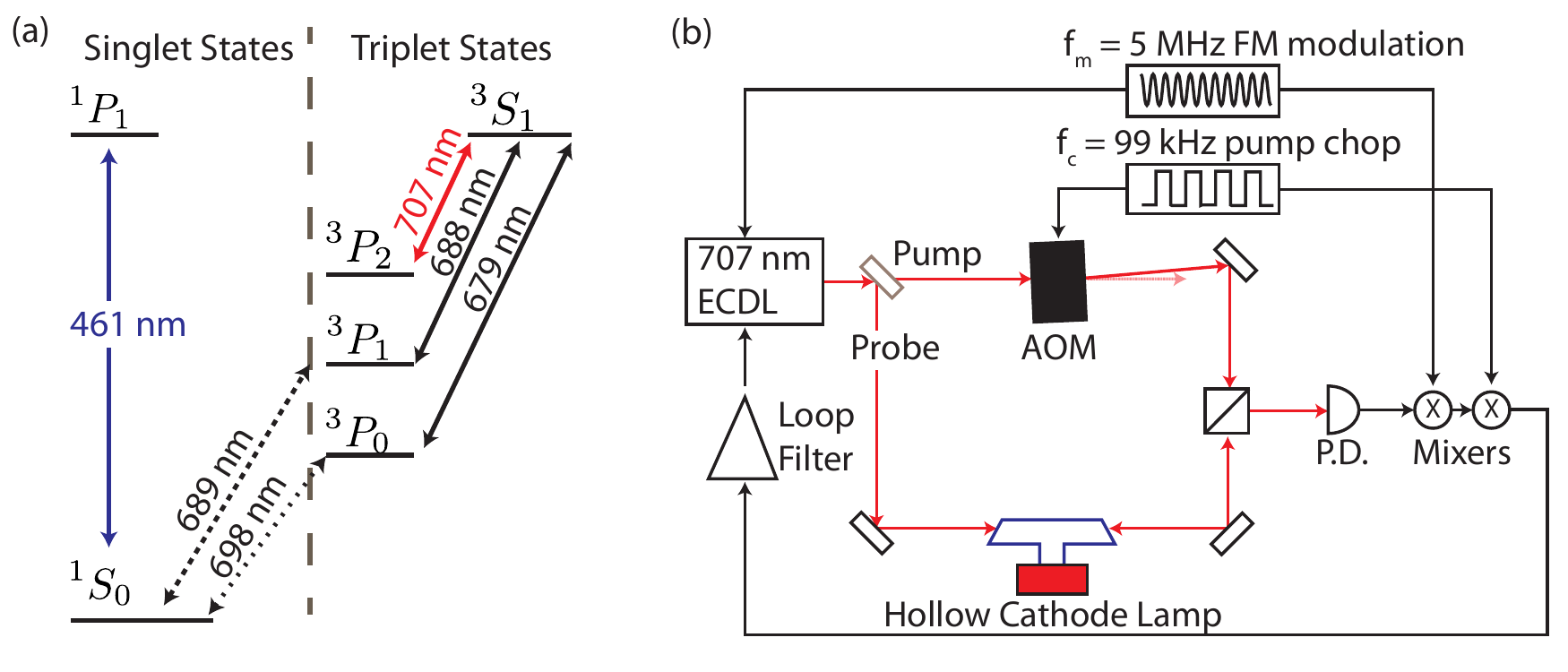}
\caption{ (a) Relevant atomic transitions in $^{88}$Sr, with approximate wavelengths noted.  The lasers described in this paper address the $^3\mathrm{P}_2$ to $^3\mathrm{S}_1$ transition near 707~nm (shown in red), and the $^1$S$_0$ to $^1$P$_1$ transition near 461~nm (shown in blue).  (b) Diagram of experimental apparatus.  The extended-cavity diode laser at 707~nm is stabilized to the $^3$P$_2$ to $^3$S$_1$ transition by Doppler-free FM spectroscopy.  Chopping of the pump beam reduces DC offsets and associated sensitivity to DC offset drifts.} 
\label{fig:Exp}
\end{figure*}

\begin{figure}
\includegraphics[width=3.5in]{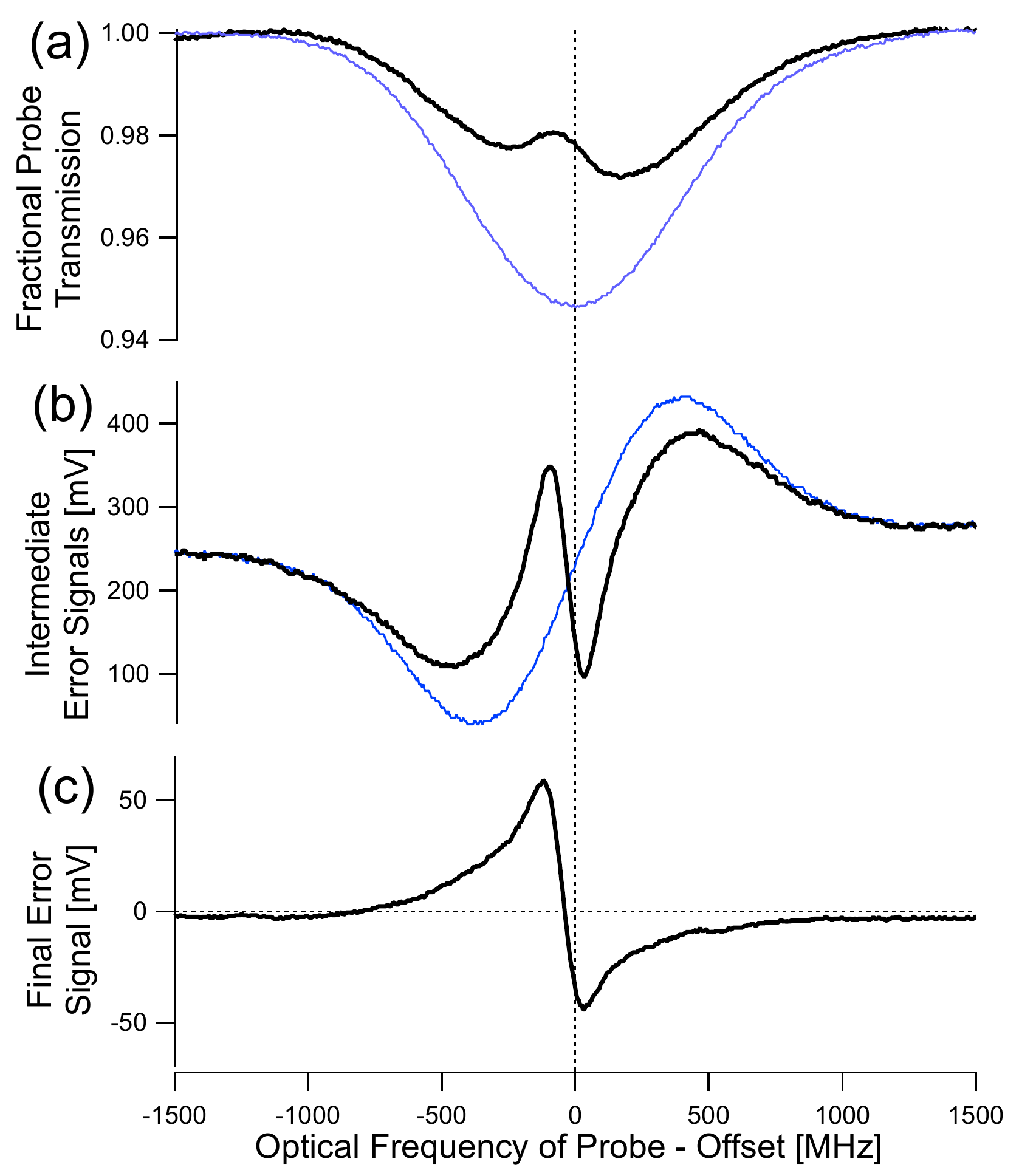}
\caption{ Transmitted power and error signals versus laser frequency.  (a)  The transmitted probe power exhibits a Doppler broadened linewidth when the pump laser is blocked (blue, thin), and a narrower Doppler free feature (black, thick)  when the counter-propagating pump is not blocked.  (b) The demodulated FM-spectroscopy signals without pump (blue thin) and with pump (black thick.)  The modulation and demodulation here is only at $f_m$ (c) The FM spectroscopy error signal after an additional chopping of the pump and demodulation stage at $f_c$.  The DC offset is significantly reduced, and the capture range is increased.  As expected, the zero of the error signal is displaced from the fitted center of the doppler broadened profile (thin blue) in (a) by $f_s/2= 40$~MHz due to the AOM's 80~MHz shift of the pump frequency.}  
\label{fig:ErrSig}
\end{figure}

Figure \ref{fig:ErrSig}a shows the fractional amount of probe power detected in transmission as the laser frequency is slowly swept through the Doppler broadened resonance, with both the pump blocked (thin blue) and the pump unblocked (thick black).  The horizontal frequency scale is calibrated by injecting some of the laser light into a Michelson interferometer and simultaneously recording the interferometer fringes with spacing of 980(9)~MHz.  The measurement bandwidth of a single trace is approximately 2~kHz, and 100 traces are averaged for display purposes.   A Gaussian fit to the transmission gives an rms 1 dimensional velocity for the atoms of 280(10)~m/s or an equivalent temperature $\mathrm{T}= 550(50)^\circ$~C.  The maximal transmission dip with no pump is roughly $5\%$.  When the pump is unblocked, a Doppler-free transmission peak appears in the center of the broader transmission dip. The peak is  about $1\%$ of the total transmission signal.

Frequency modulation spectroscopy is employed to provide a robust error signal for frequency stabilizing the laser to the Doppler-free feature.  The laser current is modulated at $f_m =5$~MHz to generate frequency modulation sidebands that appear on both the pump and probe. Figure \ref{fig:ErrSig}b shows the resulting error signals after demodulation of the photodiode signal at $f_m$ using a mixer (Minicircuits ZPRD-1).  The output of the mixer is shown here for a single sweep after filtering with a single-pole RC low pass filter with corner frequency of $f_f=1.5$~kHz  or noise equivalent bandwidth NEB = 2.4~kHz. The signal is amplified prior to demodulation by 41 dB (Minicircuits ZKL-1R5.)  The signals with (thick black) and without (thin blue) the pump are shown and are roughly the derivatives of the transmission signals in Fig.~\ref{fig:ErrSig}a.

The current modulation of the laser produces both an FM and AM response.  The AM component of the modulation produces much of the large DC offset in Fig.~\ref{fig:ErrSig}b.  This DC offset can vary in time leading to frequency errors in the lock point, for instance with slowly drifting laser cavity geometry or other optical etalons in the probe optical path.  To suppress these drifts, the pump passes through an $f_s$=80 MHz AOM and the diffracted first order (80 MHz lower in frequency than the probe) is utilized as the pump beam with all other orders blocked.  The AOM chops the pump  on and off with 50\% duty cycle at $f_c = 99.0~$kHz by switching the 80~MHz rf on and off using a fast TTL-driven switch (Minicircuits ZASWA-2-50DR.)  The final error signal used for locking the laser is shown in Fig.~\ref{fig:ErrSig}c and  is recovered by a second demodulation step at $f_c$  (using a Minicircuits ZAD-8 mixer).  The error signal shown is a single sweep filtered as before with corner frequency $f_f=1.5$~kHz.

The chopping at $f_c$ suppresses the DC offset by a factor ~100 to roughly -2~mV.  The DC offset can be converted to a frequency error of 2~MHz using the central slope of the error signal $\alpha=1.1$~MHz/mV.  Reducing the DC offset makes the lock point much more stable in time.  The peaks of the dispersive error signal are separated by 160~MHz, and the error signal maintains the correct sign for locking over a range of roughly $\pm800$~MHz.  The extended range (relative to approximately $\pm 150$~MHz  without chopping of the pump at $f_c$) allows the laser to more robustly recover from large transient perturbations such as dropping metal objects on the optical table and also facilitates the pre-alignment of the laser frequency before engaging the feedback loop.  

For applications where current modulation of the laser is unacceptable, we have used a phase modulator (AR coated lithium niobate crystal) in the probe path.  At the cost of additional complexity, this approach eliminates the 5~MHz sidebands imposed on the laser by direct current modulation and produces very similar results.  


The error signal is used in a single integrator feedback loop with unity gain frequency of approximately 1.6~kHz, roughly matched to the low pass filter used to display the error signals in Fig.~\ref{fig:ErrSig}.  The same correction signal is applied to both the laser current and a piezo adjusting the grating angle, with the relative magnitude of the feedback chosen to optimize the mode hop free tuning range of the laser (approximately 5~GHz).  

\begin{figure}
\includegraphics[width=3.5in]{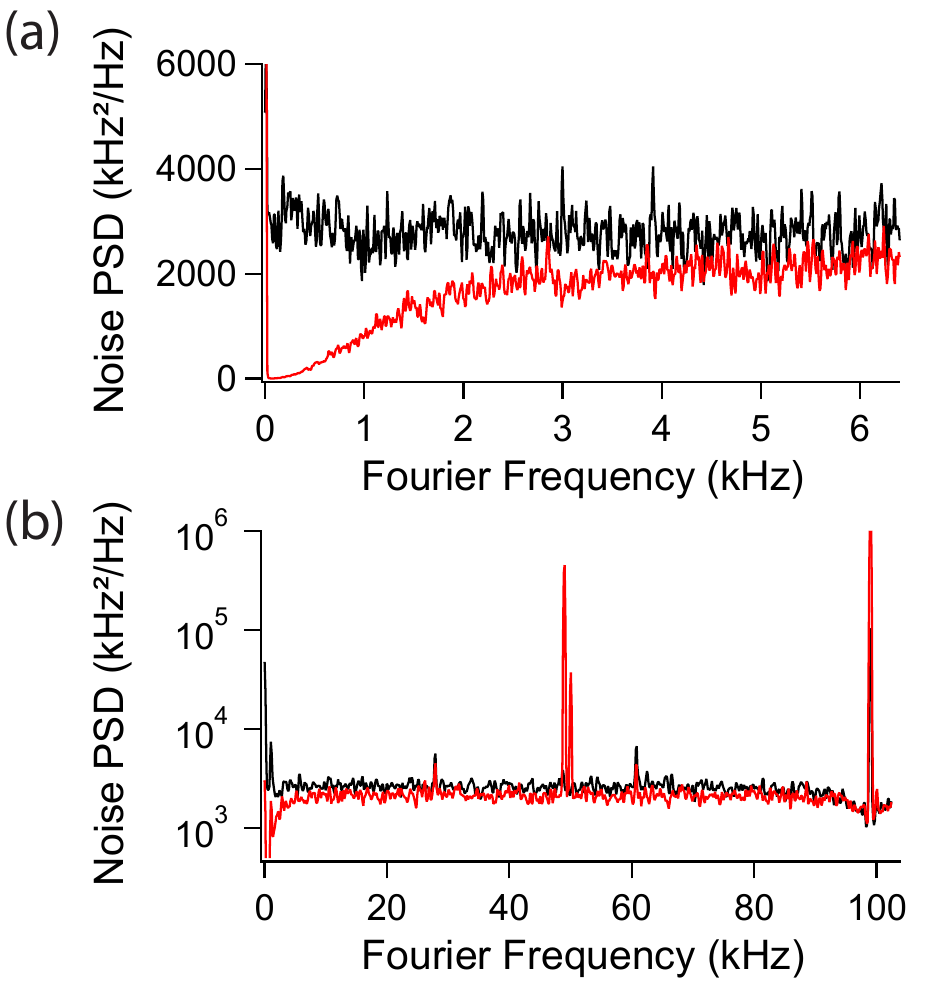}
\caption{(color online) Power spectral density of instantaneous frequency noise of the 707~nm error signal in lock (red) and out of lock and off resonance with atomic transition (black).    (a) Broadband measurement noise is imposed on the laser frequency at frequencies below $\sim1$~kHz when in lock.  This contributes 1.7~MHz rms fluctuations on the laser frequency in a 1~kHz noise equivalent bandwidth.  (b) Chopping of the pump produces a peak in the error signal spectrum at $f_c=99$~kHz. An intermodulation between the frequency modulation at $f_m$ and the chopping leads to an additional peak near 50~kHz due to an intermodulation between $f_c$ and $f_m$. The peaks are well outside of the feedback bandwidth.  } 
\label{fig:noise}
\end{figure}

Figure 3 shows the measured power spectrum of the error signal both in closed loop (red traces) and in open loop, with the laser tuned well outside of atomic resonance (black traces).  The fitted slope of the error signal (91~kHz/mV for the lithium niobate phase modulator configuration used for this data) has been used to convert voltage noise to instantaneous frequency noise.   

There are several features to note in the closed loop error signal's power spectrum.  First, the broadband noise is mainly due to photon shot noise in the probe (80\% of the noise variance) with a small additional contribution from the 2.3~pA$/\sqrt{\mathrm{Hz}}$ input current noise of the AD8015 transimpedance amplifier at $f_m$ (20\% of the noise variance).  Below the unity gain frequency, this noise is imposed on the laser's actual optical frequency, as can be seen in Figure 3a from the dip in the flat noise floor at low frequencies.  We estimate that the broadband noise generates fluctuations in the laser frequency of 1.7~MHz rms at 1~kHz noise equivalent bandwidth.  This is much less than the excited state linewidth $\Gamma= 13.2$~MHz. The servo bandwidth is chosen as a compromise between suppressing ambient acoustic noise and actually imposing additional frequency noise on the laser.

We also note that the noise spectra exhibit large spikes near~50~kHz and at 99~kHz.  The 99~kHz frequency component arises directly from the modulation and demodulation at $f_c$.  The 50~kHz spikes arise from an intermodulation product between the 5~MHz modulation and the 99~kHz chopping.  For small changes in $f_m$ and $f_c$ the frequency of these peaks change in the relative ratio of  1:1 and roughly 50:1 respectively, such that the 50~kHz spike's frequency changes rapidly with small changes in $f_c$.  The value of $f_c$ was tuned to maximize the frequency of the intermodulation signals so that they are well above the unity gain frequency of the laser frequency servo and therefore would not be imposed onto the laser.  Using phase coherent synthesizers for $f_m$ and $f_c$, it should be possible to phase coherently cancel this intermodulation product.  However, assuming a single integrator feedback loop, these peaks generate an rms frequency deviation of the laser of 400~kHz rms and the present approach is sufficient.

We find that the use of the 707~nm repump laser increases the steady-state $^1$S$_0$ MOT population by a factor of four. The use of a second repump laser at 679~nm in addition to the 707~nm repump allows an additional factor of ten population increase.

\begin{figure}
\includegraphics[width=3.5in]{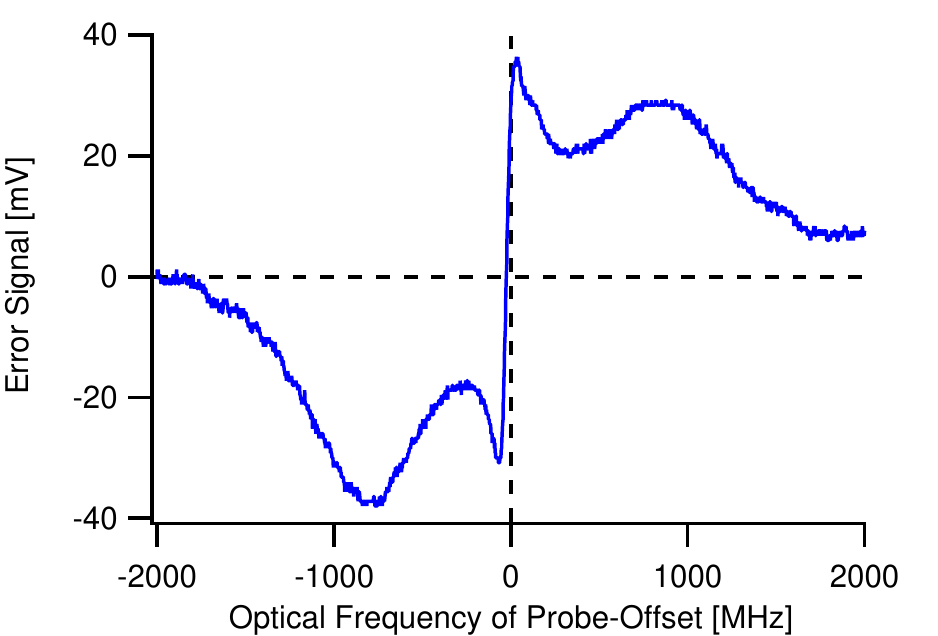}
\caption{Error signal derived from polarization spectroscopy of the $^1$S$_0$ to $^1$P$_1$ transition at 461~nm, using the same hollow cathode as used for the 707~nm lock.  } 
\label{fig:461}
\end{figure}

The same hollow cathode lamp is shared with a polarization spectroscopy setup that stabilizes a 461~nm laser to the $^1$S$_0$ to $^1$P$_1$ transition (actual wavelength of 460.862~nm) in a manner very similar to that described in \cite{Yoshio13}.  The 707~nm and 461~nm beams are overlapped on dichroic mirrors but are spatially offset from each other such that they pass through different parts of the hollow cathode lamp.  The error signal obtained from the 461~nm spectroscopy is shown in Figure \ref{fig:461}.  The 461~nm pump and probe beam powers are 1~mW and 50~$\mu$W, respectively.  The waist size for both beams is 0.2~mm.  As with the 707~nm spectroscopy setup, the pump beam is chopped at 99~kHz to provide supression of background offsets.  The central slope of the error signal is 800~MHz/V.  

The extra lobes on the error signal at roughly the width of the Doppler-broadened absorption feature are unexpected.  One possible explaination is that they are due to pump photons scattered by the high optical depth (96\% absorption on resonance) gas within the hollow cathode lamp.  

The error signal has a flat 5.6~kHz/$\sqrt{\mathrm{Hz}}$ noise floor.  The noise has roughly equal contributions of photon shot noise and white noise on the balanced photodiode detector.  This noise contributes 180~kHz of added noise to the laser for feedback of 1~kHz noise equivalent bandwidth. This is much less than the 32~MHz half-linewidth of the $^1$S$_0$ to $^1$P$_1$ transition.

In conclusion, we present a simple and robust method for stabilizing a laser to the 707.202~nm transition in strontium, with future applications for a broad range of precision and many-body physics in laser-cooled strontium atoms.  We have also used a separate laser to successfully observe error signals for the $^{3}\mathrm{P}_1$ and $^{3}\mathrm{P}_0$ to $^{3}\mathrm{S}_1$ transitions at 688 and 679 nm respectively, although the signal to noise in the present configuration precluded stabilization to the error signal with $ >10$~Hz bandwidth.   Future optimizations may increase the signal to noise for locking, perhaps by utilizing a hybrid lock utilizing the locked 707~nm light as a counter propagating pump for probe light at 688 or 679~nm.

\section{Acknowledgments}
We gratefully acknowledge Gretchen Campbell who related her group's observation of the 707~nm transition in a hollow cathode lamp. Part numbers are given as technical information only, and do not represent endorsement by NIST. This material is based upon work supported by the National Science Foundation under Grant Number 1125844 Physics Frontier Center, NIST, DARPA QUASAR, and ARO.

\bibliography{707spectroscopybib}

\end{document}